\documentclass[conference,compsoc]{IEEEtran}
\usepackage{times}
\usepackage{latexsym}
\usepackage{graphicx,subfigure}
\usepackage{epstopdf}
\usepackage{amsmath}
\usepackage{amssymb}
\usepackage{breqn, algorithm, algpseudocode}
\usepackage{multirow}
\ifCLASSOPTIONcompsoc
  \usepackage[nocompress]{cite}
\else
  \usepackage{cite}
\fi
\ifCLASSINFOpdf
\else
\fi
\begin{document}
%
\title{Network-based Distance Metric with Application to Discover Disease Subtypes in Cancer}
%
%
%
%

\author{\IEEEauthorblockN{{Jipeng Qiang$^{1}$, Wei Ding$^1$}, John Quackenbush$^2$, Ping Chen$^1$}
	\IEEEauthorblockA{$^1$Department of Computer Science, University of Massachussets Boston \\
	$^2$Biostatistics and Computational Biology, Dana-Farber Cancer Institute
		}
	}

%
%

\maketitle
%




\begin{abstract}
While we once thought of cancer as single monolithic diseases affecting a specific organ site, we now understand that there are many subtypes of cancer defined by unique patterns of gene mutations. These gene mutational data, which can be more reliably obtained than gene expression data, help to determine how the subtypes develop, evolve, and respond to therapies. Different from dense continuous-value gene expression data, which most existing cancer subtype discovery algorithms use, somatic mutational data are extremely sparse and heterogeneous, because there are less than 0.5\% mutated genes in discrete value 1/0 out of 20,000 human protein-coding genes, and identical mutated genes are rarely shared by cancer patients. 

    Our focus is to search for cancer subtypes from extremely sparse and high dimensional gene mutational data in discrete 1 and 0 values using unsupervised learning. We propose a new network-based distance metric. We project cancer patients' mutational profile into their gene network structure and measure the distance between two patients using the similarity between genes and between the gene vertexes of the patients in the network. Experimental results in synthetic data and real-world data show that our approach outperforms the top competitors in cancer subtype discovery. Furthermore, our approach can identify cancer subtypes that cannot be detected by other clustering algorithms in real cancer data. 

\end{abstract}

\IEEEpeerreviewmaketitle

\section{Introduction}

%
%
%
%
Identifying cancer subtypes is essential for a wide range of applications including better understanding the biological complexity of the disease and developing targeted, precision medicine therapeutic interventions \cite{network:comprehensive, network:cancer}. Subtype discovery is a fundamental yet unsolved problem in cancer analysis as the presence of multiple subtypes can confound many analyses \cite{panagiotis:gene-expression, panagiotis:gene}.

Gene mutational data, which can be more reliably obtained than gene expression data, help to determine how the subtypes develop, evolve, and respond to therapies \cite{olivier2011somatic, pirazzoli2014acquired}. Different from dense continuous-value gene expression data, which most existing cancer subtype discovery algorithms use, somatic mutational data are extremely sparse and heterogeneous, because there are less than 0.5\% mutated genes out of 20,000 human protein-coding genes, and identical mutated genes are rarely shared by cancer patients \cite{hofree2013network}. Clustering algorithms are often used for cancer subtype discovery. The major barrier for clustering algorithms is how to battle with extremely sparse and high dimensional gene mutational data in discrete 1 and 0 values. 

We propose a new network-based distance metric to take advantage of the prior knowledge of gene regulator networks. We project cancer patients' mutational profile into their gene network structure and measure the distance between two patients using the similarity between genes and between the gene vertexes of the patients in the network. We introduce a novel metric to measure gene similarity that fully utilizes network structures, and patient gene mutational profiles are optimally aligned based on network structures.  Experimental results show that our approach outperforms the top competitors in cancer subtype discovery using a comprehensive set of evaluation metrics. Furthermore, our approach can identify cancer subtypes with biological significance that cannot be detected by other clustering algorithms in real cancer data.

Thus, our main contributions are as follows:

(1) Network-based distance metric: we propose a novel distance metric to measure similarity between cancer patients using gene regulator networks.

(2) Effectiveness: Our approach outperforms state-of-the-art algorithms in discovering cancer subtypes, and detects biological significant cancer subtype that cannot be identified by the top competitors from real cancer data.

Furthermore, our network-based distance metric can be easily incorporated into any clustering algorithm with application to data whose attributes have network structures. 

Reproducibility: Our code is open-sources at https://github.com/qiang2100/NetAP.

\section{A Novel Network-based Distance Metric}

We introduce a new approach of gene similarity and patient similarity to compute the similarity between two patient profiles by incorporating gene interaction information in gene networks.\\ 
\textbf{Overview of our network-based distance metric}. Even if two patients do not have any mutations in common, they are likely to belong to the same cluster when their mutations reside in close network regions. The gene interaciton network includes the relationship between genes, where each node represents one gene. Take the example of three patients with three mutated genes: p1=(8,10,11), p2=(1,3,7), p3=(5,22,29), and one correspondent gene interaction network. Because these patients do not share any common mutated genes, the similarity among the three patients is zero using traditional distance metrics. However, if we measure the mutated genes residing in a gene interaction network after projecting three patients into the gene network (see Figure 1), it is clear that p1 and p2 are more likely to belong to one cluster (same cancer subtype) than p1 and p3, because the mutated genes of p1 are biologically similar to the mutated genes of p2 than p3. 

\begin{figure}
	\centering
	\includegraphics[width=88mm]{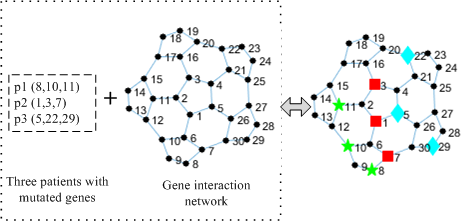}
	\caption{An illustration of three patients (p1, p2, and p3) with mutated genes and their correspondent gene network. The total number of vertices in this network is 30. The mutated genes of p1, p2 and p3 are marked in green stars, red squares, and cyan diamonds, respectively. Although these three patients do not share any common mutated genes, p1 and p2 are more likely to belong to one subtype than p1 and p3, because the mutated genes of p1 are closer to the mutated genes of p2 than p3.}
\end{figure}

Through independently projecting each patient into the gene network, we transform the similarity between two patients to the similarity of two sub-networks.  This approach has two advantages: (1) it reduces the dimensionality since we only need consider these genes from the network; (2) two patients sharing similar network regions will have higher similarity than two patients not sharing similar network regions. Our distance metric calculation includes two steps:

1) We need to compute the similarity between genes using the gene network.
 
2) After projecting each patient into the gene network, we need to match the vertices of two patients, and apply our distance metric to compute the similarity between two patients. 
 \\
\textbf{Gene Similarity}. We first present how to compute the similarity between individual node(gene) pairs according to the gene network. Let $sim(u,v)$ represent the similarity between vertice \textit{u} and vertice \textit{v}. Since two patients are more likely to belong to the same cluster when their mutations share the similar network regions, the closer the distance between two vertices in the network, the higher similarity they have. The similarity between vertices can be calculated with two cases: the similarity of a vertice with itself and the similarity of different vertices. The pseudocode of computing the similarities between genes is described in Algorithm 1.

For the first case, we first assign 1 (lines 2 to 4) as a initial value to the similarity of one vertex with itself. In a typical approach, the similarity between a gene and itself should be 1. However, in our new proposed distance metric, the similarity between the same gene from two different cancer patients is calculated based on the netwrok that the gene resides. We will first compute the similarity between other vertices, and then modify the value based on their corresponding neighbors in order to distinguish the influence of different vertices (see discussion for Equations 2, 3, and 4 later in this section). 

If \textit{u} and \textit{v} have a direct connection in the gene network, they are neighbors. The similarity between two vertices is similar to the degree of closeness of relationship between two vertices. For two different vertices, the similarity between two vertices should become smaller as the distance increases. The closeness between two vertices are determined by their number of neighbors. For two different vertices ($u$ and $v$), similarity can be calculated by finding the greatest similar path from one vertex to another vertex. 

At first, for two vertices that are adjacent, its initial vaule are set using traditional approach (lines 5 to 7),
\begin{equation}sim(u,v) = \dfrac{\mid edge(u)\mid \cap \mid edge(v)\mid}{ \mid edge(u) \mid \cup \mid edge(v)\mid}\end{equation}
 where $edge(u)$ represents all edges of vertex \textit{u}, and $\mid edge(u)\mid$ represents the number of $u$'s edges. 

We will update the similarity between two vertices by finding the greatest similar path using other vertices as intermediate points along the way. We define a function $similarPath(i, j, k)$ that returns the greatest similar path from \textit{i} to \textit{j} using vertices only from the set \{1,2,...,\textit{k}\} as intermediate points along the way. After defining this function, our aim is to find the greatest similar path from each \textit{i} to each \textit{j} using only vertices from 1 to \textit{k} + 1 (lines 8 to 16). The greatest similar path of each pair of vertices must be either (1) a path that only uses vertices in the set \{1,2,...,\textit{k}\}, or (2) a path that goes from \textit{i} to \textit{k}+1 and then from \textit{k}+1 to \textit{j}. Consequently, we can define $sim(i, j, k + 1)$ recursively: the base case is

 \begin{equation}similarPath(i,j,0) = sim(i,j)\end{equation}

 and the recursive case is

 \begin{dmath}similarPath(i,j,k+1) = max(similarPath(i,j,k), similarPath(i,k+1,k) \times similarPath(k+1,j,k))\end{dmath}
 
Equation 3 ensures that the similarity between pair $i$ and $j$ is always the greatest similar path.  

The formula is similar to Floyd's algorithm that can be used for finding shortest paths in a weighted network \cite{rivest1990introduction, rosen2011discrete}. In Floyd's algorithm, it adds all the weights along the path. In our method, we multiply them using Equation 3. The strategy computes \textit{similarPath}(\textit{i}, \textit{j}, \textit{k}) for all (\textit{i}, \textit{j}) pairs for \textit{k} = 1, then \textit{k} = 2, until \textit{k} = \textit{m}, and we can find the similar path for all (\textit{i}, \textit{j}) pairs using any intermediate vertices. Finally, similarity between a vertex \textit{g} and itself can be updated as the sum of its similarity from all its neighbor's vertices in gene network (lines 17 to 19),
 \begin{equation}sim(g,g) = \sum_{j \in neigh(g)}^{}sim(g,j)\end{equation}
 Here, $neigh(g)$ represents all neighbors of vertice \textit{g}. The underlying principle of Equation 4 is that the similarity between a gene and itself in a densely connected network is greater than a gene in a loosely connected network.

\begin{algorithm}
\textbf{Algorithm 1} Gene similarity using gene network
\begin{algorithmic}[1]

	\State Let \textit{sim} be a \textit{m}$\times$\textit{m} matrix that  are initialized to zero
	\For{\texttt{each vertex \textit{g}}}
	\State \texttt{$sim(g,g) \gets 1$}
	\EndFor
	\For{\texttt{each edge (u,v)}}
		\State \texttt{$sim(u,v) \gets $}  Equation 1
		\EndFor
	\For{\texttt{k from 1 to m}}
		\For{\texttt{i from 1 to m}}
			\For{\texttt{j from 1 to m}}
			\If {$sim(i,j)\leq sim(i,k) \times sim(k,j)$}
			\State $sim(i,j)\gets sim(i,k) \times sim(k,j)$
			\EndIf
			\EndFor
		\EndFor
	\EndFor
	\For{\texttt{each vertex \textit{g}}}
	\State \texttt{$sim(g,g) \gets \sum_{j \in neigh(g)}^{}sim(g,j)$}
	\EndFor
\end{algorithmic}
\end{algorithm}

The complexity of computing the similarity between genes is O$(m^{3}$), where \textit{m} is the number of genes in gene network. Note that in order to reduce the computational cost of the above algorithm, a threshold can be set for $sim(i,k)$ (e.g., 1e-6 for this work) so that the dissimilarity between two genes is filtered out and the algorithm can be accelerated greatly. \\
\textbf{Patient Similarity}. After projecting each patient into the gene network, we compute the similarity between two patients through their own mutated genes's distance in the network. To distinguish from the other metrics, we refer to the distance metric as gene aligning's similarity (GAS). Let $r_{i}$ and $r_{j}$ be the representation of patient \textit{i} and patient \textit{j}. We can think of $r_{i}$ as a set of vertices in gene network, where $r_{i,g}=1/(\mid r_{i}\mid )$ is the weight of the vertex \textit{g} of patient \textit{i}. Here, $\mid r_{i} \mid$ is the number of the vertices of patient $r_{i}$ in gene network. The task of patient similarity is how to optimally align the genes of patient $i$ and patient $j$ to properly calculate the maximal similarity.

First, we measure the alignment between vertices $r_{i}$ and $r_{j}$ by calculating the weight of vertices $r_{i}$ align to the vertices $r_{j}$ through the gene network. Let \textit{T} $\in \mathbb{R}^{m \times m}$ be a align matrix, where \textit{T}$_{u,v}$ represents how much the weight of vertex \textit{u} of $r_{i}$ matches to vertex \textit{v} of $r_{j}$, and $m$ is the number of genes in the gene network. Further, to match all weights of $r_{i}$ into $r_{j}$, the entire outgoing weight from vertex \textit{u} equals $r_{i,u}$, namely $\sum_{v}^{}T_{u,v} = r_{i,u}$. Correspondingly, the amount of incoming weight to vertex \textit{v} must equal $r_{j,v}$, namely, $\sum_{u}^{}T_{u,v} = r_{j,v}$. At last, we can define the similarity of two patients as the maximum cumulative cost required to align from all vertices of one patient to the other patient, namely,  $\sum_{u,v}^{}T_{u,v} sim(u,v)$. Given the constraints, the similarity between two patients can be solved using the following linear programming,
 
 \begin{displaymath}\max_{T \geq 0} \sum_{u,v}^{m}T_{u,v} sim(u,v)\end{displaymath}  
 \begin{equation}such \; that: \sum_{v}^{m}T_{u,v} = r_{i,u} \quad \forall{u} \in \{1,2,...,m\}\end{equation}
 \begin{displaymath}\quad \quad \quad \quad \quad \sum_{u}^{m}T_{u,v} = r_{j,v} \quad \forall{v} \in \{1,2,...,m\}\end{displaymath}
 
 The above optimization is a special case of the Earth Mover's Distance \cite{rubner1998metric,wolsey2014integer}, a well-known transportation problem for which specialized solvers have been developed \cite{ling2007efficient, pele2009fast}. The best average time complexity of solving the GAS problem is O($m^{3}$log${m}$), where \textit{m} is the number of all genes in the gene network \cite{pele2009fast}. To speed up the optimization problem, we relax the GAS optimization problem and remove one of the two constraints. Consequently, the optimization becomes,

\begin{equation}
 \max_{T \geq 0} \sum_{u,v}^{m}T_{u,v} d(u,v) \quad  s.t. \sum_{v}^{m}T_{u,v} = r_{i,u} \forall{u} \in \{1,2,...,m\} 
\end{equation}

The optimal solution is the probability of each gene in a patient is aligned to the most similar gene in the other patient. The time complexity of GAS can be reduced to O($m$log${m}$). 
  
Recall the earlier example of three patients in Figure 1. The weight of each gene of each patient is $\frac{1}{3}$. For patient p1 and p2, the weight of gene 11 aligns to gene 3, the weight of gene 10 aligns to gene 1, and the weight of gene 8 aligns to gene 7. We note that GAS agrees with our intuition, and ''aligns'' genes to nearby genes. Consequently, the similarity of p1 and p2 (0.11) is significantly bigger than the similarity of p1 and p3 (0.00694), although all of them do not share one common mutated gene. The result meets our assumption that p1 and p2 are more likely to belong to one subtype than p1 and p3. \\
\textbf{Cancer subtype discovery}. The similarity matrix of patients is used on cancer subtype discovery via Affinity Propagation \cite{frey2007clustering}. Affinity Propagation is a clustering algorithm that takes as input measures of similarity between pairs of texts and simultaneously considers all data points as potential exemplars.  The whole method is named as Network-based Affinity Propagation (NetAP). The framework of NetAP is shown in Figure 2.

\begin{figure}
	\centering
	\includegraphics[width=90mm]{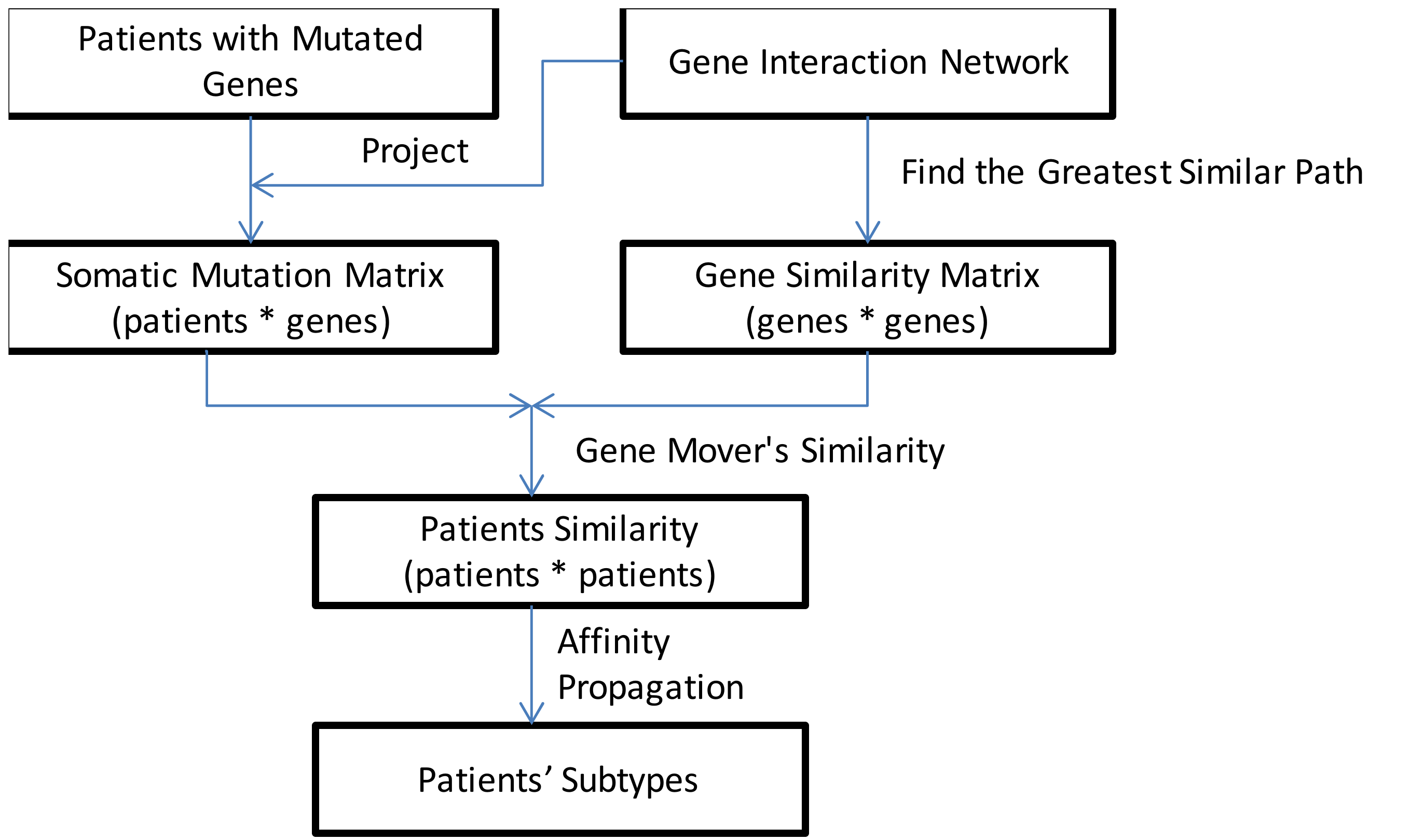}
	\caption{ Framework of Network-based Affinity Propagation.}
\end{figure}

\section{Experiment}

The task of clustering cancer patients using tumor mutation information  is difficult. A real-world cancer dataset typeically has hundreds of samples, but the number of gene mutation features can be well above 15,000 as shown in Table 1. Cancer is a complex disease. Two cancer patients of the same cancer subtype may not share any common mutated genes. Therefore, many clustering methods cannot achieve good results if they calculate two samples' distance directly using the gene mutation feature. Cancer has highly heterogeneous causes, and it is difficult to find a clear group of genes to determine subtypes. In our empirical study, we observe that AP is the strongest baseline clustering algorithm even though it does not use  gene network. A possible explanation is that the power of belief-propagation can better tune the centers of each cluster. Hence, in our experiments we chose AP to integrate with gene interaction networks for optimal performance. We evaluate our NetAP algorithm using synthetic data and real-world data with different focuses:

(1)	\textbf{Evaluation using Synthetic Data}. How accurately does NetAP detect cancer subtypes with respect to various gene network structures? Does NetAP outperform the state-of-the-art algorithms used for cancer subtype discovery?

(2)	\textbf{Performance using Real-World Data}. Does NetAP detect cancer subtypes that are clinically meaningful? What is the impact of different gene networks on performance? Can NetAP identify cancer subtypes that cannot be detected by other clustering algorithms?

We implemented NetAP in Matlab\footnote{the source code can be downloaded at https://github.com/qiang2100/NetAP}. All experiments were conducted on a Windows machine with an Intel 437 2.9 GHz CPU and 8GB memory. Table 1 and Table 2 show the details of the real-world Uterine and Lung cancer datasets and three gene interaction networks we used in our experiments. 

\subsection{Dataset Information and Experiment Setup}

\textbf{Synthetic Data}. To test the accuracy of our method, we built a synthetic dataset to mirror the biological characteristics of cancer and investigate the effectiveness of the incorporation of gene interaction networks. We randomly select four subnetwork modules from the Search Tool for the Retrieval of Interacting Genes/Proteins (STRING) gene interaction network \cite{STRING}, and yield 200 cancer patient samples with 50 samples per cancer subtype. For each sample, we reassign a fraction of mutations to fall within its network modules. Simultaneously, given an overlapping rate $r$ (0.05 and 0.1), each sample has a probability $r$ to select mutations from other network modules. Here, the overlapping rate 0.05 or 0.1 indicates how strongly the network structure is embedded in the data. Higher overlapping rate means a larger number of genes are shared among different subtypes.

\textbf{Real-world Data}. High-grade uterine endometrial carcinoma and lung adenocarcinoma somatic mutation data were collected from the The Cancer Genome Atlas (TCGA) data portal\footnote{https://tcga-data.nci.nih.gov/tcga/}. Only mutation data generated using the high-quality Illumina GAIIx platfrom were saved for the following analysis, and patients with less 10 mutations were removed for fair comparison with \cite{hofree2013network}. Patient mutation profiles are constructed as binary vectors such that a bit is set to 1 if the gene corresponding to that position in the vector is mutated in that patient. We follow the same somatic mutation data processing procedure as \cite{hofree2013network, network:cancer}.

\begin{table}
	\centering
	\caption{Summary of Uterine and Lung cancers }
	\#: the number of patients, SIZE: the number of genes, AVG: the average mutated genes of each patient
	\begin{tabular}{|c|c|c|c|} \hline
	Dataset& \# & SIZE & AVG\\\hline
		Uterine  & 248 & 17968 & 612.96\\ \hline
		Lung & 304  & 15967 & 326.83   \\ 
		\hline\end{tabular}
	
\end{table}

\textbf{Evaluation Metrics}: The clustering results on real-world data are evaluated using histological types provided by the TCGA data. Five metrics are used to measure the clustering performance: Normalized Mutual Information (NMI), Rand Index (RI), Adjusted Rand Index (AR), Chi-square test and P-Value, chosen based on established evaluation criteria in the literature. NMI, RI and AR are widely used to evaluate performance of clustering algorithms in data mining and machine learning \cite{manning2008introduction, wu2008top}. Chi-square test (Chi-Square) and P-Value are mostly used in statistics and bioinformatics \cite{agresti1996introduction}.  For NMI, RI, AR, and Chi-square, a larger score indicates better clustering performance. For p-value, a small p-value represents good clustering quality.

\begin{itemize}
\item{Normalized Mutual Information (NMI) is a clustering validation metric that effectively measures the amount of statistical information shared by the predicted cluster assignments and the ground truth, independent of the absolute cluster label values.  Two patients are assigned to the same cluster if and only if they are similar, thus clustering can be viewed as a series of pair-wise decisions.}

\item{Rand Index (RI) measures the percentage of clustering decisions that are correct. Rand Index can be adjusted for the chance grouping of elements, which will result in one of its variants called Adjusted Rand Index (AR). AR has a value between 0 and 1, and RI can have negative values.}

\item{Chi-squared test is used to determine whether there is a significant difference between the expected clusters and the observed clusters.}

\item{P-Value can determine how significant clustering results are by performing a hypothesis test commonly used in statistics.}
\end{itemize}

\textbf{Existing State-of-the-art Methods for comparison}: We compared our NetAP algorithm with Nonnegative Matrix Factorization (NMF) \cite{lee2001algorithms}, Latent Dirichlet Allocation (LDA) \cite{blei2003latent}, Affinity Propagation (AP) \cite{frey2007clustering}, and Network-based stratification(NBS) \cite{hofree2013network}. NMF, LDA, and NBS are the leading clustering algorithms on cancer subtype discovery. AP has shown better performance on computational biology tasks than K-means. We use the open-source MATLAB implementation\footnote{https://sites.google.com/site/nmftool/} for NMF based on Euclidean distance. LDA based on gibbs sampling is chosen as comparison\footnote{http://psiexp.ss.uci.edu/research/programs_data/toolbox.htm} \cite{griffiths2004finding} with parameters $\alpha$=0.1 and $\beta$=0.1. For AP, we use the ''apcluster'' package in R\footnote{https://cran.r-project.org/web/packages/apcluster/index.html}. Based on empirical observation, Peason correlation coefficient is chosen as the distance metric. The source code of NBS is provided in Hofree et al. \cite{hofree2013network}. We set parameter $\lambda$=0.9 for AP and NetAP.

\textbf{Gene Interaction Network}: To evaluate the impact of different gene interaction network, three major gene interaction databases are used: PathwayCommons \cite{pathwayCommons}, STRING \cite{STRING} and HumanNet \cite{HumanNet}. PathwayCommons\footnote{www.pathwaycommons.org/pc/} includes gene interaction information extracted from multiple gene interaction databases, and its focus is on physical protein-protein interactions. We excluded all non-human genes and interactions from the PathwayCommons network in our experiments. STRING\footnote{www.string-db.org/} collects protein-protein interactions from expression data analysis and medical literature using text mining methods. HumanNet\footnote{www.functionalnet.org/humannet/} is built by a modified Bayesian integration from multiple organisms. Only the top 10\% interactions of STRING and HumanNet are used in our experiments to reduce noise. Table 2 summarizes the number of genes and interactions, and the numbers in parentheses are specific for our experiments. 

\begin{table}
\centering
\caption{Summary of gene interaction networks }
\begin{tabular}{|c|c|c|} \hline
	 &Nodes&Edges \\\hline
	 PathwayCommons  & 14,355 (2814) & 507,757 (33,757)\\ \hline
	STRING & 16,569 (12,233) & 1,638,830 (164,034)  \\ \hline
	HumanNet & 16,243 (7,949) & 476,399 (47,641)  \\ 
	\hline\end{tabular}

\end{table}

\subsection{Evaluation on Synthetic Data}

In the 5 sub-figures of Figure 3, we demonstrate that NetAP can effectively detect cancer subtypes with respect to weak and strong gene network structure. The comparison is done with NMF, AP, LDA and NBS on synthetic data using five evaluation metrics (NMF, RI, AR, Chi-square, P-value). We run each algorithm 20 times, and all  results  of these five metrics are the average value of 20 times per experimental setting.  NMF, LDA and AP, which are the three algorithms that do not utilize gene interaction networks, in general, have worse performance compared to NetAP and NBS, which are the two algorithms using gene network information.  

Furthermore, NetAP works robustly and is  better than  NBS.  By increasing the rate of overlapping from 0.05 to 0.1, we find that the performance of all methods is worse  due to more noise in each clustering (the clustering membership is less certain due to the increased overlap). NetAP remains the best out of all these methods.

\begin{figure}
{
	\includegraphics[width=.5\linewidth]{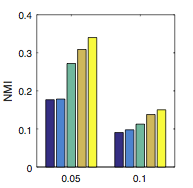}\hfill%
	\includegraphics[width=.5\linewidth]{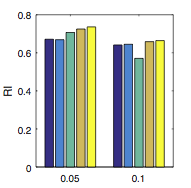}%
}
{
	\includegraphics[width=.5\linewidth]{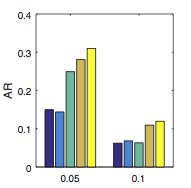}\hfill%
	\includegraphics[width=.5\linewidth]{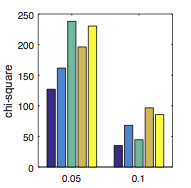}%
}
{
	\includegraphics[width=.5\linewidth]{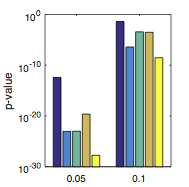}%
	\includegraphics[width=.4\linewidth]{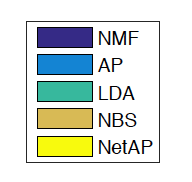}%
}
\caption{Performance of all methods with two different overlapping rates(0.05 and 0.1)  on synthetic data). NetAP is our proposed method. For P-value metric, the smaller the better. For other metrics, the larger the better. }
\end{figure}

With Lung caner data, NetAP achieves the best performance compared to all other methods. However, although NBS still outperforms NMF, it has similar performance with AP that does not take advantage of gene network structure. We suspect that NMF-based methods (NBS is based on NMF) still struggle with extremely sparse data such as somatic mutation data that has lots of 0s and few 1s, even though that incorporating network information can help to alleviate the spareness problem to a certain degree.

By analyzing the results of two real-world cancer datasets, we found that AR of LDA and NMF are very close to zero that is barely better than random assignment.  In Table 3, we showed that the well-established clustering algorithm SparseNMF (NMF using L1 regularization), Kmeans, PAM, Hierarchical clustering algorithms have almost identical or worse performance than random assignment.  

In our work, we assume that cancer patients belonging to one subtype are more likely to share a similar network subregion. Network based NBS (also NMF based) achieves better results than NMF, and NetAP outperforms all other methods that we compared against in real-world data. Because NMF and its variations do not work very well due to sparse and heterogeneous characteristics of somatic mutation feature space.

In summary, we can conclude that NetAP is the most appropriate clustering algorithm for clustering gene mutations.  Because NBS and NetAP are the only two algorithms using gene network, we will compare them in more details using the other two gene networks.

\begin{figure}
\centering

 	\includegraphics[width=.8\linewidth]{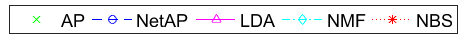}%
 
{
	\includegraphics[width=.5\linewidth]{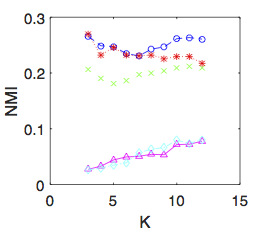}\hfill%
	\includegraphics[width=.5\linewidth]{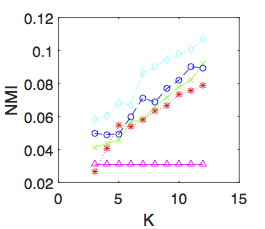}%
}

{
	\includegraphics[width=.5\linewidth]{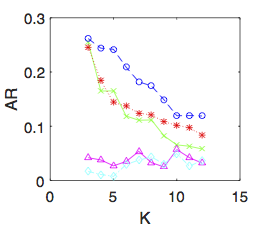}\hfill%
	\includegraphics[width=.5\linewidth]{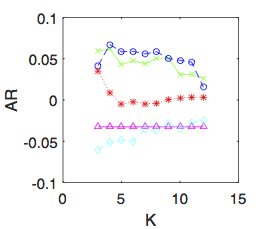}%
}
{
	\includegraphics[width=.5\linewidth]{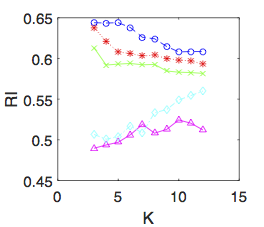}\hfill%
	\includegraphics[width=.5\linewidth]{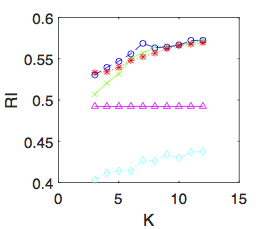}%
}
{
	\includegraphics[width=.5\linewidth]{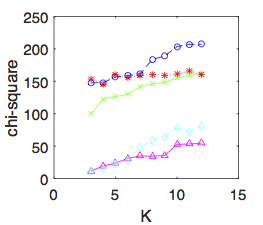}\hfill
	\includegraphics[width=.5\linewidth]{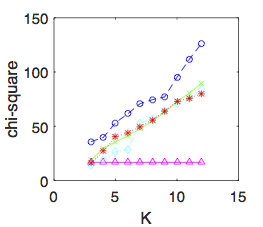}
}
{
	\includegraphics[width=.5\linewidth]{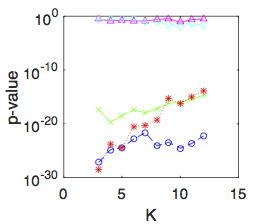}\hfill%
	\includegraphics[width=.5\linewidth]{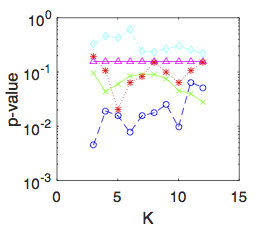}%
}

\caption{Performance of NetAP compared to NMF, LDA, AP, and NBS with different values of \textit{K} using NMI, Rand index, Adjusted Rand Index, Chi-square and P-value metrics on uterine and lung Cancer. NetAP is our proposed method. For P-value, the smaller the better. For others, the larger the better. }
\end{figure}

We chose four existing methods (NMF, LDA, AP, NBS) in the previous experiment. To fully assess the effectiveness of our method, we have conducted more experiment to compare with other clustering algorithms (SparseNMF \cite{hoyer2004non}, Kmeans \cite{wu2008top}, PAM \cite{han2011data} and Hierarchical \cite{murtagh2014ward}). For conciseness we only show the results using NMI metric. ''Random'' refers to the result by random drawing. The performance of these existing methods is very similar to the results of ''Random'', which means all these methods are not effective for somatic mutation stratification due to extreme sparseness of somatic mutations. Therefore, incorporating the knowledge of gene network to reduce sparseness is very important for identifying subtypes from somatic mutation data.

\begin{table}
	\centering
	\caption{Performance of NetAP and other clustering methods on uterine cancer using NMI. }
	\begin{tabular}{|c|c|c|c|} \hline
		K &3&4&5 \\\hline
		Random & 0.0138 & 0.0199& 0.0245\\ \hline
		SparseNMF & 0.0247 & 0.0141 & 0.0198 \\ \hline
		Kmeans & 0.0332 & 0.1035 & 0.1040 \\ \hline
		PAM & 0.0136 & 0.0708  & 0.1073\\ \hline
		Hierarchical & 0.0107 & 0.0708 & 0.1073\\ \hline
		NetAP & 0.2659 &  0.2488 & 0.2473 \\
		\hline\end{tabular}
\end{table}

\subsubsection{Impact of Gene Networks}
Figure 5 shows the performance of NetAP and NBS on the uterine cancer dataset,  incorporating the other two networks (STRING and HumanNet) with different numbers of subtypes (\textit{K}=3, 4, ..., 12) using five metrics (NMF, RI, AR, Chi-square, P-value). Clearly NetAP works better than NBS on these two gene interaction networks in general, except on AR and RI metrics using the STRING network. Especially, when increasing the number of subtypes \textit{K}, NetAP can achieve better results than NBS. The experimental results give further evidence that our method is more robust for subtype identification.  

\begin{figure}

\centering

   \includegraphics[width=.5\linewidth]{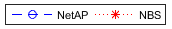}%
{
	\includegraphics[width=.5\linewidth]{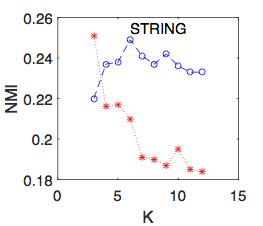}\hfill%
    \includegraphics[width=.5\linewidth]{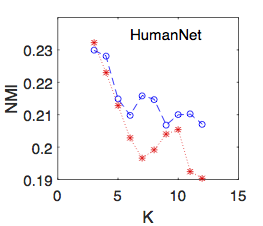}%
}

{
	\includegraphics[width=.5\linewidth]{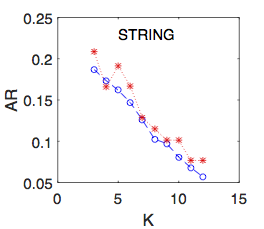}\hfill%
    \includegraphics[width=.5\linewidth]{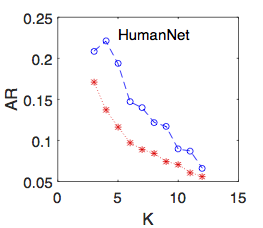}%
}
{
	\includegraphics[width=.5\linewidth]{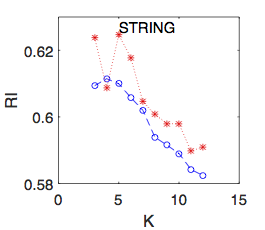}\hfill%
    \includegraphics[width=.5\linewidth]{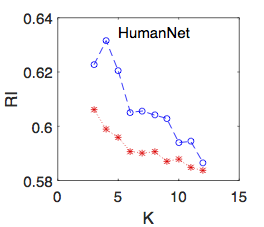}%
}
 {
 	\includegraphics[width=.5\linewidth]{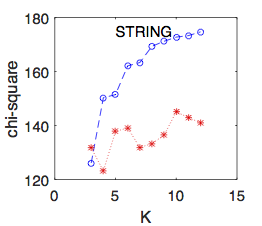}\hfill
 	\includegraphics[width=.5\linewidth]{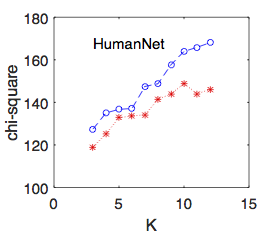}
 }
 {
 	\includegraphics[width=.5\linewidth]{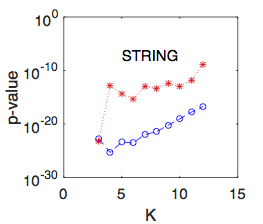}\hfill%
 	\includegraphics[width=.5\linewidth]{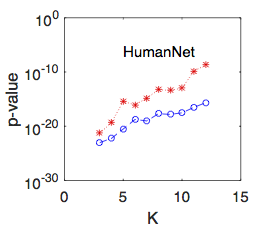}%
 }

\caption{Performance of NBS and NetAP on the other two human networks (STRING and HumanNet) with respect to different values of \textit{K}. For P-value, the smaller the better. For others, the larger the better.  }
\end{figure}

As NetAP is naturally dependent on gene interaction networks, we examine how different gene networks affect the quality of NetAP with NMI metric. We chose the following three gene networks: PathwayCommons, STRING and HumanNet. Figure 6 shows the results of NetAP  with different gene networks on uterine cancer dataset. When varying the subtypes from 3 to 12, NetAP using PathwayCommons or STRING performs superior to NetAP using HumanNet. Additionally, NetAP using PathwayCommons outperforms NetAP using STRING. In conclusion, the performance of NetAP will vary when it incorporates different gene networks, because different network structures. The new finding indicates that PathwayCommons can provide strong genetic trait on cancer subtype discovery.

\begin{figure}
	\centering
	\includegraphics[height=60mm]{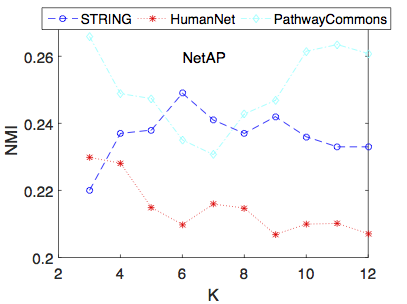}
	\caption{Performance of NetAP with three gene networks (PathwayCommons, STRING and HumanNet) using NMI on Uterine cancer. }
\end{figure}

\subsubsection{Identified Subtypes}

To assess the biological significance of the identified subtypes, we examine whether they correlate with observed clinical data. Figure 7 shows the results of NMF, LDA, AP, NetAP and NBS with the recorded subtypes on a histological basis. We can see that NetAP subtypes are more closely associated with the recorded subtypes on the histological basis than other algorithms. NMF and LDA cannot separate any type of ''serous adenocarcinoma type'' and ''endometrioid type'' from the data set. NBS can only extract one subtype ''serous adenocarcinoma type''. NetAP and AP can separate the two subtypes ''serous adenocarcinoma type'' and ''endometrioid type''. Furthermore, NetAP has higher accuracy than AP.

\begin{figure}
	\centering
	\includegraphics[height=42mm]{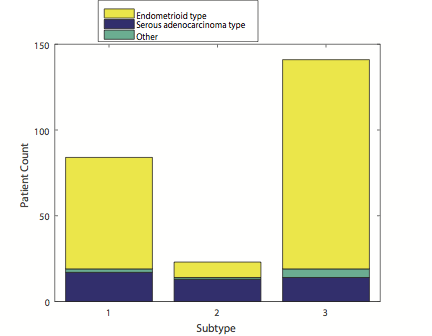} \\{{\tiny (a) NMF}}\\
	\includegraphics[height=42mm]{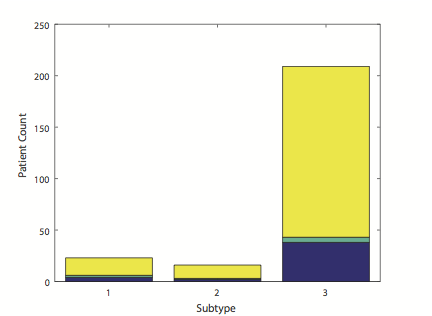}\\{{\tiny (b) LDA}}\\
	\includegraphics[height=42mm]{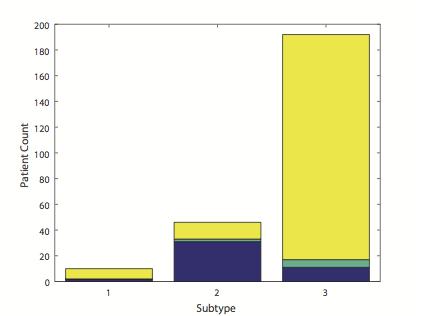}\\{{\tiny (c) AP}}\\
	\includegraphics[height=42mm]{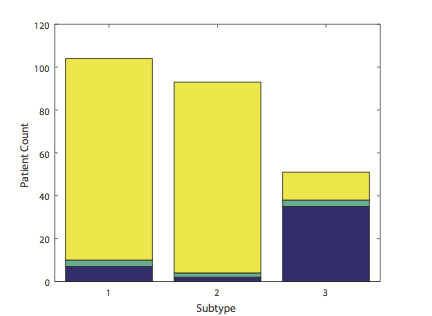}\\{{\tiny (d) NBS}}\\
	\includegraphics[height=42mm]{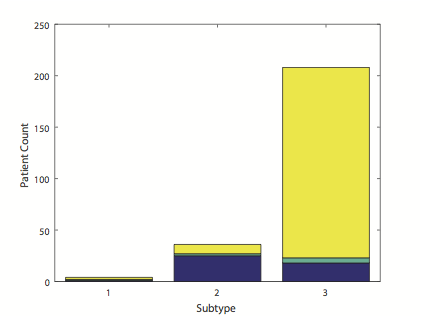}\\{{\tiny (e) NetAP}}\\
	\caption{Summary of Histological types for each subtype on Uterine Cancer.  }
\end{figure}



\section{Related Work}

Clustering is difficult because it belongs to unsupervised learning and there does not always exist an unambiguous membership for a data point. Many clustering algorithms have been proposed to discover hidden structures for a variety of applications \cite{brunet2004metagenes, wang2015extended, wu2008top}. Popular clustering algorithms include K-means \cite{wu2008top}, its variation PAM\cite{han2011data}, Hierarchical clustering \cite{murtagh2014ward}, DBSCAN \cite{ester1996density}, Latent dirichlet allocation (LDA) \cite{blei2003latent}, Hierarchical Dirichlet process \cite{teh2006hierarchical}, Principal component analysis(PCA) \cite{han2011data} and Affinity Propagation (AP) \cite{frey2007clustering}. Based on whether data elements can belong to one cluster or more than one cluster, clustering algorithms can be categorized as hard clustering or soft clustering. In hard clustering, data is divided into distinct clusters, where each data element belongs to exactly one cluster, e.g. K-means algorithm \cite{han2011data} and Affinity Propagation (AP)\cite{frey2007clustering}. In soft clustering (also referred to as fuzz clustering), data elements can belong to more than one cluster, and associated with each element is a set of membership levels, e.g. Non-negative Matrix Factorization (NMF) \cite{brunet2004metagenes, lee2001algorithms} and Latent dirichlet allocation (LDA) \cite{blei2003latent, griffiths2004finding}.

If we focus on clustering algorithms to stratify sparse and heterogeneous somatic mutational profiles, perhaps the most popular approach for subtype discovery is NMF, which does not require any priori knowledge of the expected number of subtypes or the associated mutational patterns \cite{brunet2004metagenes}. NMF aims to find two non-negative matrices whose product provides a good approximation to the original matrix. One of its drawbacks is that it does not always result in meaningful parts-based clustering representations. Several researchers addressed this problem through incorporating sparseness constraint (sparse NMF) on one or both non-negative matrices \cite{hoyer2004non, kim2015kdd}. In addition, NetNMF, one of NMF variants, encoded the geometrical structure in the data to regularize one of the two non-negative matrices \cite{cai2008non}.  NMF has been applied to recover meaningful biological information from cancer-related microarray data without supervision \cite{brunet2004metagenes, wang2013non}. However even NMF with sparseness constraints cannot effectively stratify somatic mutation data because of its extremely sparseness. Since there are a variety of gene interaction networks, Network-based stratification (NBS) \cite{hofree2013network} adopted NetNMF algorithm for handling somatic mutational profiles. So far, NBS is the only method to stratify patients in an unsupervised fashion from somatic mutation data. However, its performance still needs significant improvement for practical clinical application. With high-dimensional sparse data the other natural choice is to learn a low-dimensional representation using techniques like Latent Semantic Indexing (LSI)\cite{deerwester1990indexing} or Latent Dirichlet Allocation (LDA)\cite{blei2003latent}, however these methods often fail to show any meaningful improvement on somatic mutation clustering.  

\section{Conclusion}
In this paper, we study the critical problem of stratifying tumor mutation profiles into subtypes through incorporating gene interaction information. We present a new framework called Network-based Affinity Propagation (NetAP) to cluster tumor mutations with gene networks. To use the knowledge of the gene network, we project patient profiles into a gene interaction network and develop a new distance metric called gene aligning's similarity to compute the similarity between patients. We demonstrate the effectiveness and efficiency of our approach on synthetic and uterine adenocarcinoma datasets along with three popular gene networks using five different metrics. In future, we plan to integrate multiple layers of information beyond somatic mutations (e.g. CNVs, transcriptome, etc.) into our method for better subtype identification.

\ifCLASSOPTIONcompsoc
  \section*{Acknowledgments}
\else
  \section*{Acknowledgment}
\fi

The work was supported in part by the NVIDIA foundation's Compute the Cure program and the State Scholarship Fund of the China Scholarship Council. 

\ifCLASSOPTIONcaptionsoff
  \newpage
\fi



%

\bibliography{PatientSum} 
\bibliographystyle{ieeetr}

%




\end{document}